\documentclass[12pt]{article}
\usepackage{amsfonts,pifont}
\usepackage{amsmath,chemarrow}
\usepackage{amssymb}
\usepackage{amstext}
\usepackage{amsfonts}
\usepackage{graphicx}
\usepackage{indentfirst}
\usepackage{hyperref,comment}
\makeatletter
\def\ExtendSymbol#1#2#3#4#5{\ext@arrow 0099{\arrowfill@#1#2#3}{#4}{#5}}
\def\RightExtendSymbol#1#2#3#4#5{\ext@arrow 0359{\arrowfill@#1#2#3}{#4}{#5}}
\def\LeftExtendSymbol#1#2#3#4#5{\ext@arrow 6095{\arrowfill@#1#2#3}{#4}{#5}}
\makeatother

\begin{document}
\baselineskip 20pt

\title{Interpretation of $Z_c(4025)$ as the Hidden Charm Tetraquark States via QCD Sum Rules}
\author{Cong-Feng Qiao$^{a,b}$\footnote{qiaocf@ucas.ac.cn}\; and
Liang Tang$^{a}$\footnote{tangl@ucas.ac.cn} \\[0.5cm]
$a)$ School of Physics, Graduate University of Chinese Academy of Sciences\\ YuQuan Road 19A, 100049, Beijing, China\\[0.2cm]
$b)$ CAS Center for Excellence in Particle Physics, Beijing, China\\[0.2cm]}


\date{}
\maketitle

\begin{abstract}

By using QCD Sum Rules, we found that the charged hidden charm tetraquark $[c u][\bar{c} \bar{d}]$ states with $ J^P = 1^-$ and $ 2^+$, which are possible quantum numbers of the newly observed charmonium-like resonance $Z_c(4025)$, have masses of $m_{1^-}^c = (4.54 \pm 0.20) \, \text{GeV}$ and $m_{2^+}^c = (4.04 \pm 0.19) \, \text{GeV}$. The contributions up to dimension eight in the Operator Product Expansion (OPE) were taken into account in the calculation. The tetraquark mass of $J^{P} = 2^{+}$ state was consistent with the experimental data of $Z_c(4025)$, suggesting the $Z_c(4025)$ state possessing the quantum number of $J^P = 2^+$. Extending to the b-quark sector, the corresponding tetraquark masses $m_{1^-}^b = (10.97 \pm 0.25) \, \text{GeV}$ and $m_{2^+}^b = (10.35 \pm 0.25) \, \text{GeV}$ were obtained, which are testable in future B-factories.

\end{abstract}

\section{Introduction}

A charged charmonium-like state $Z_c(4025)$ has been reported
by BESIII in the process $e^+ e^- \rightarrow (D^* \bar{D}^*)^\pm
\pi^\mp$ \cite{Ablikim:2013nra}. Its mass and width are respectively
$(4026.3 \pm 2.6 \pm 3.7) \; \text{MeV}$ and $(24.8 \pm 5.7 \pm 7.7)
\; \text{MeV}$. The mass and decay modes imply that it is a charged
hidden charm state, which is similar to $Z_c(3900)$
\cite{Ablikim:2013mio, Liu:2013dau, Xiao:2013iha}. In addition, the BESIII Collaboration recently announced another structure called $Z_c(3885)$~\cite{Ablikim:2013xfr} in the invariant mass spectrum of $(D \bar{D}^*)^{\pm}$. The existing analyses~\cite{Aceti:2014uea, Braaten:2014qka} favor this state with the quantum number of $J^P = 1^+$. Were the $Z_c(3885)$ of the same origin as the $Z_c(3900)$, the quantum numbers of the $Z_c(3900)$ would be $1^+$. However, the quantum number of $Z_c(4025)$ so far is not well determined.

This paper utilize standard techniques of the QCD Sum Rules
\cite{Shifman, Reinders:1984sr, Narison:1989aq, P.Col} to
investigate the masses of charged hidden charm tetraquark states
with two quantum numbers, {\it i.e.} $J^P = 1^-$ and $2^+$. The
hidden charm tetraquark state for $Z_c(4025)$ is investigated
through examination of experimental data. Utilizing the QCD Sum
Rules, the hidden charm tetraquark states with various quantum
numbers have been investigated in Refs.\cite{Qiao:2013raa,
Matheus:2006xi, Chen:2010ze, Chen:2012pe, Dias:2013xfa}, yielding
significant conclusions. For $J^{P} = 1^-$, an unstable mass sum
rules was obtained \cite{Chen:2010ze}, where the interpolating
current was consistent with the $1^-$ current. However,  stable
results were extracted in Refs.\cite{Albuquerque:2008up, Albuquerque:2012zy}. This paper
reanalyzes this case by adding several new ingredients
\cite{Qiao:2013raa} and performing moderate criteria
\cite{Matheus:2006xi}, to determine the available threshold
parameter $\sqrt{s_0}$ and the Borel window $M_B^2$.

It should be noted that, very recently, the BESIII Collaboration has observed a charged charmonium-like resonance in the processes $e^+ e^- \to (h_c \pi^\pm) \pi^\mp$, named $Z_c(4020)$ \cite{Ablikim:2013wzq}. So far it is still too early to tell whether the $Z_c(4025)$ and the $Z_c(4020)$ are the same origin or not \cite{Yuan:2013lma}, though many theoretical investigations have already done \cite{Braaten:2014qka, Guo:2013sya, Khemchandani:2013iwa, Aceti:2014kja, Wang:2013bxa, Chen:2013wva, Wang:2013llv}. Among them, Braaten {\it et al.} interpreted the exotic states as Born-Oppenheimer tetraquarks which are $[c \bar{c}][q\bar{q}]$ (color octet-octet) states~\cite{Braaten:2014qka}; Guo {\it et al.} suggested the $Z_c(4025)$ as a $D^* \bar{D}^*$ virtual state~\cite{Guo:2013sya}; Khemchandani {\it et al.} discussed the possibilities of the $Z_c(4025)$ being a $1^+$ or $2^+$ $D^* \bar{D}^*$ bound state in the framework of QCD Sum Rules~\cite{Khemchandani:2013iwa}; and Aceti {\it et al.} also argued that the $Z_c(4025)$ is a $2^+$ state, but being a $D^* \bar{D}^*$ bound state~\cite{Aceti:2014kja}.

In Sec.II, various essential formulae are presented. Numerical
analysis and mass extraction are shown in Sec.III, with conclusions
given in Sec.IV.

\section{Formalism}

The QCD Sum Rules begin with the two-point correlation functions :
\begin{eqnarray}
\Pi_{\mu \nu}(q) & = & i \int d^4 x e^{i q \cdot x} \langle 0 | T
\big{\{} j_\mu(x) j^\dagger_\nu(0) \big{\}} | 0 \rangle \; , \\
\Pi_{\mu \nu, \, \alpha \beta}(q) & = & i \int d^4 x e^{i q \cdot x}
\langle 0 | T \big{\{} j_{\mu \nu}(x) j^\dagger_{\alpha \beta}(0)
\big{\}} | 0 \rangle \; .
\end{eqnarray}

The interpolating currents of $1^-$ and $2^+$ hidden charm
tetraquark states are respectively constructed as:
\begin{eqnarray}
j_\mu^{1^-}(x) \!\!\! & = & \!\!\! \frac{i \epsilon_{a b c}
\epsilon_{d e c}}{\sqrt{2}} \big[ \left( u^T_a(x) C \gamma_5 c_b(x)
\right) \left( \bar{d}_d \gamma_\mu \gamma_5 C \bar{c}_e^T \right) -
\left( u^T_a(x) C \gamma_\mu \gamma_5 c_b(x) \right) \left(
\bar{d}_d \gamma_5 C \bar{c}_e^T \right) \big] \; , \\ j_{\mu
\nu}^{2^+}(x) \!\!\! & = & \!\!\! \frac{i \epsilon_{a b c}
\epsilon_{d e c}}{\sqrt{2}} \big[ \left( u^T_a(x) C \gamma_\mu
c_b(x) \right) \left( \bar{d}_d \gamma_\nu C \bar{c}_e^T \right) -
\left( u^T_a(x) C \gamma_\nu c_b(x) \right) \left( \bar{d}_d
\gamma_\mu C \bar{c}_e^T \right) \big] \;, \label{current}
\end{eqnarray}
where, $a$, $b$, $c$, $\cdots$, are color indices, and $C$
represents the charge conjugation matrix.

For $j^{1^-}_\mu(x)$, the correlation function has the following
Lorentz covariance form:
\begin{eqnarray}
\Pi_{\mu\nu}(q) = - \left(g_{\mu \nu} -\frac{q_\mu
q_\nu}{q^2}\right) \Pi_1(q^2) + \frac{q_\mu q_\nu}{q^2} \Pi_0(q^2)
\; . \label{function1}
\end{eqnarray}
where the subscripts 1 and 0 respectively denote the quantum numbers
of the spin 1 and 0 mesons.

The two-point function of the current $j^{2^+}_{\mu \nu}(x)$ has the
following Lorentz form \cite{Chen:2011qu}:
\begin{eqnarray}
\Pi_{\mu \nu, \, \alpha \beta}(q)  & = &  T_{\mu \nu, \, \alpha
\beta} \Pi_2 (q^2) + \cdots\; .
\end{eqnarray}
Here $\Pi_2(q^2)$ is the part of $\Pi_{\mu \nu, \, \alpha \beta}(q)$
which exclusively projects onto the $2^+$ state, and $T_{\mu \nu, \,
\alpha \beta}$ is the unique Lorentz tensor of the fourth rank
constructed from $g_{\mu \nu}$ and $q_\mu$:
\begin{eqnarray}
T_{\mu \nu, \, \alpha \beta}=\frac{1}{2} \big[g^t_{\mu \alpha}(q)
g^t_{\nu \beta}(q) + g^t_{\mu \beta}(q) g^t_{\nu \alpha}(q) -
\frac{2}{3} g^t_{\mu \nu}(q) g^t_{\alpha \beta}(q) \big] \;,
\end{eqnarray}
which satisfies the following desired properties:
\begin{eqnarray}
& & T_{\mu \nu, \, \alpha \beta} = T_{\alpha \beta, \, \mu \nu}, \;
\; q^\mu
T_{\mu \nu, \, \alpha \beta}  = 0 \;, \nonumber \\
& & g_t^{\mu \nu}(q) T_{\mu \nu, \, \alpha \beta} = 0 \; ,
\end{eqnarray}
where $g^t_{\mu \nu}(q) = (g_{\mu \nu}- q_\mu q_\nu/q^2)$.

On the phenomenological side, after separating out the ground state
contribution from the pole term of the $\Pi_{i}(q^2)$, where $i = 1$
or $2$, the correlation function is expressed as a dispersion
integral over a physical regime,
\begin{eqnarray}
\Pi_i(q^2) & = & \frac{\lambda_{i}^{c \; 2}}{m_{i}^{c \; 2} - q^2} +
\frac{1}{\pi} \int_{s_0}^\infty d s \frac{\rho_{i}^h(s)}{s - q^2}
\;, \label{spectraldensity}
\end{eqnarray}
where $m_i^c$, $\lambda_i^c$ and $\rho_i^h(s)$ respectively
represent the mass, decay constant, and spectral density of the
tetraquark state. Here $s_0$ is the threshold of higher excited
states and continuum states, and $\lambda_i^c$ is defined in
Refs.\cite{Qiao:2013raa, Chen:2011qu}. It is worth to note that the tetraquark state defined here couples to the four-quark interpolating field, which has a special structure differing from the $D^* \bar{D^*}$ configuration. And hence the $D^* \bar{D}^*$ threshold effect is neglected here.

On the OPE side of the $\Pi_i(q^2)$, the correlation function is
expressed as a dispersion relation:
\begin{eqnarray}
\Pi_i^{OPE}(q^2) = \int_{4 m_c^2}^{\infty} d s
\frac{\rho_i^{OPE}(s)}{s - q^2} \; ,
\end{eqnarray}
where $\rho_i^{OPE}(s) = \text{Im} [\Pi_i^{OPE}(s)] / \pi$, and is
expressed as:
\begin{eqnarray}
\rho_i^{OPE}(s) & = & \rho_i^{pert}(s) + \rho_i^{\langle \bar{q} q
\rangle}(s) + \rho_i^{\langle g_s^2 G^2 \rangle}(s) +
\rho_i^{\langle g_s \bar{q} \sigma \cdot G q \rangle}(s) +
\rho_i^{\langle \bar{q} q \rangle^2}(s) + \rho_i^{\langle g_s^3 G^3
\rangle}(s) \nonumber \\ & + & \rho_i^{\langle g_s \bar{q} \sigma
\cdot G q \rangle \langle \bar{q} q \rangle}(s) + \cdots \; ,
\end{eqnarray}
where the ``$\cdots$" stands for other higher dimension condensates
omitted in our work.

To evaluate the spectral density of the OPE side, the ``full"
propagators $S^q_{i j}(x)$ and $S^Q_{i j}(p)$ of a light quark
($q=u$, $d$ or $s$) and a heavy quark ($Q=c$ or $b$) are
respectively written with the vacuum condensates clearly displayed
\cite{Reinders:1984sr}.
\begin{eqnarray}
S^q_{i j}(x) \! \! & = & \! \! \frac{i \delta_{i j} \hat{x}}{2 \pi^2
x^4} - \frac{m_q \delta_{i j}}{4 \pi^2 x^2} - \frac{i g_s t^a_{i j}
G^a_{\kappa \lambda}}{32 \pi^2 x^2}(\sigma^{\kappa \lambda} \hat{x}
+ \hat{x} \sigma^{\kappa \lambda}) + \frac{i\delta_{i j}
\hat{x}}{48} m_q \langle \bar{q}q \rangle  - \frac{\delta_{i j}
\langle \bar{q} q \rangle}{12} \nonumber \\ \! \! \! & - & \! \! \!
\frac{\delta_{i j} \langle g_s \bar{q} \sigma G q \rangle x^2}{192}
- \frac{t^a_{i j} \sigma^{\kappa^\prime \lambda^\prime}}{192}
\langle
g_s \bar{q} \sigma \cdot G^\prime q \rangle + \cdots \;,
\end{eqnarray}
\begin{eqnarray}
S^Q_{i j}(p) \! \! & = & \! \! \int \frac{d^4 p}{(2 \pi)^4} e^{-i p
\cdot x} \bigg\{ \frac{i}{\hat{p} - m_Q}\delta_{i j} - \frac{i}{4}
g_s (t^c)_{i j} G^c_{\kappa \lambda} \frac{1}{(p^2 - m_Q^2)^2}
\nonumber \\ \! \! & \times & \! \! [\sigma^{\kappa \lambda}
(\hat{p} + m_Q) + (\hat{p} + m_Q) \sigma^{\kappa \lambda}] +
\frac{i}{12} g_s^2 \delta_{i j} G^a_{\alpha \beta} G^a_{\alpha
\beta} m_Q \frac{p^2 + m_Q \hat{p}}{(p^2 - m_Q^2)^4} \nonumber \\
\! \! & + & \! \! \frac{i \delta_{i j}}{48} \bigg[ \frac{(\hat{p} +
m_Q) [\hat{p} (p^2 - 3 m_Q^2) + 2 m_Q (2 p^2 - m_Q^2)] (\hat{p} +
m_Q)}{(p^2 - m_Q^2)^6} \bigg] \langle g_s^3 G^3 \rangle + \cdots
\bigg\} \; ,
\end{eqnarray}
where the Lorentz indices $\kappa^\prime$ and $\lambda^\prime$
correspond to the indices of an outer gluon field from another
propagator, and $G^\prime$ represents the outer gluon field
\cite{Albuquerque:2013ija}.

Using the techniques of Refs.\cite{Matheus:2006xi, Qiao:2013raa} the
spectral density $\rho_i^{OPE}(s)$ was calculated up to dimension
eight at the leading order in $\alpha_s$.

The $1^{-}$ tetraquark state spectral density of the OPE side are
given as:
\begin{eqnarray}
\rho_{1}^{pert} (s) \! \! \! & = & \! \! \! \frac{1}{3 \times 2^9
\pi^6} \int_{\alpha_{min}}^{\alpha_{max}} \frac{d
\alpha}{\alpha^3}\int_{\beta_{min}}^{1-\alpha} \frac{d
\beta}{\beta^3} \bigg[\frac{3}{2} \left( 1-(\alpha + \beta)^2
\right) {\cal F} \nonumber \\ \!\!\! & + & \!\!\! 6 (m_u + m_d) m_c
(1 - \alpha - \beta)^2 + m_c^2(1
- \alpha - \beta)^3 \bigg] {\cal F}^3 \; , \\
\rho_{1}^{\langle \bar{q} q \rangle}(s) \! \! \! & = & \! \! \!
\frac{m_c \langle \bar{q} q\rangle}{2^4 \pi^4 }
\int_{\alpha_{min}}^{\alpha_{max}}\frac{d \alpha}{\alpha} \bigg[
\frac{(m_u + m_d)}{2^2 (1 - \alpha)}{\cal H}^2 + m_c
\int_{\beta_{min}}^{1 - \alpha} \frac{d \beta}{\beta} \big[ \frac{(1
- \alpha - \beta) }{\alpha} {\cal F} \nonumber \\ \!\!\! & - &
\!\!\! \frac{(m_u + m_d) m_c}{4}(\alpha + \beta + 3)\big]
{\cal F} \bigg]  \; , \\
\rho_{1}^{\langle g_s^2 G^2\rangle}(s) \! \! \! & = & \! \! \!
\frac{\langle g_s^2 G^2 \rangle}{3 \times 2^{10} \pi^6}
\int_{\alpha_{min}}^{\alpha_{max}}\frac{d \alpha}{\alpha}
\int_{\beta_{min}}^{1 - \alpha} \frac{d \beta}{\beta^2} \bigg[ \big[
(2\alpha + 2\beta -1) {\cal F} \nonumber \\ \!\!\! & + & \!\!\!
\frac{m_c^2}{\alpha}(1 - \alpha - \beta)^2 (17\alpha - 17\beta - 5)
+ \frac{3m_c^2}{\beta} (1 - \alpha -\beta) \nonumber \\ \!\!\! &
\times & \!\!\! (1 - 2\beta + (\alpha + \beta)(3\alpha + \beta))
\big] {\cal F} + \frac{m_c^4 \alpha }{\beta}
(1 - \alpha - \beta)^3 \bigg] \; , \\
\rho_{1}^{\langle g_s \bar{q} \sigma \cdot G q \rangle} (s) \!\!\! &
= & \!\!\! \frac{m_c \langle g_s \bar{q} \sigma \cdot G q
\rangle}{2^5 \pi^4} \int_{\alpha_{min}}^{\alpha_{max}} d \alpha
\int_{\beta_{min}}^{1 - \alpha} \frac{d \beta}{\beta} \bigg[ 1 +
\frac{1}{24 \alpha \beta} \left(\alpha +13\alpha^2 + 19\alpha \beta
\right. \nonumber \\ \!\!\! & + & \!\!\! \left. 6 \beta^2 -6 \beta
\right) \bigg] {\cal F} \; , \nonumber \\
\rho_{1^-}^{\langle \bar{q} q \rangle^2} (s) \!\!\! & = & \!\!\!
\frac{\langle \bar{q} q \rangle^2}{24\pi^2}
\int_{\alpha_{min}}^{\alpha_{max}} d \alpha
\big[{\cal H} - 2m_c^2 \big] \; , \\
\rho_{1}^{\langle g_s^3 G^3 \rangle} (s) \!\!\! & = & \frac{\langle
g_s^3 G^3 \rangle}{3 \times 2^{11}
\pi^6}\int_{\alpha_{min}}^{\alpha_{max}} d \alpha
\int_{\beta_{min}}^{1 - \alpha} \frac{d \beta}{\beta^3} \big[
\left(1 - \left(\alpha + \beta \right)^2 \right){\cal F} \nonumber
\\ \!\!\! & + & \!\!\! m_c^2 \left( 1 - \alpha - \beta \right)
\left( \left(1 - \alpha \right)^2 + 4 \alpha \beta + 3 \beta^2
\right) \big] \; ,
\end{eqnarray}
\begin{eqnarray}
\rho_{1}^{\langle \bar{q} \sigma \cdot G q
\rangle \langle \bar{q} q \rangle} (s) \!\!\! & = & \!\!\!
\frac{\langle \bar{q} \sigma \cdot G q \rangle \langle \bar{q} q
\rangle}{3 \times 2^{3} \pi^2} \int_{\alpha_{min}}^{\alpha_{max}} d
\alpha \bigg[ \alpha \left( \frac{3}{4} - \alpha \right) \bigg] \; ,\\
\Pi_{1}^{\langle g_s^3 G^3 \rangle} (M_B^2) \! \! \! & = & \! \! \!
-\frac{m_c^4 \langle g_s^3 G^3 \rangle}{3^2 \times 2^{11}} \int_0^1
d \alpha \int_0^{1 - \alpha} \frac{d \beta}{\beta^4} (1 - \alpha -
\beta)^4 \text{Exp}[-\frac{m_c^2(\alpha + \beta)}{\alpha \beta
M_B^2}] \; ,
\\ \Pi_{1}^{\langle g_s \bar{q} \sigma \cdot G q \rangle \langle
\bar{q} q \rangle} (M_B^2) \! \! \! & = & \! \! \! \frac{m_c^2
\langle g_s \bar{q} \sigma \cdot G q \rangle \langle \bar{q} q
\rangle}{2^4 \pi^2}\int^1_0 d \alpha \bigg[ 1 + \frac{1}{3 (\alpha
-1 )} + \frac{2 m_c^2}{3 \alpha (1 - \alpha) M_B^2} \bigg] \nonumber
\\ \!\!\! & \times & \!\!\! \text{Exp}[- \frac{m_c^2}{(1 - \alpha)\alpha
M_B^2}] \; , \; \; \; \; \; \; \; \; \; \; \label{rhoO8}
\end{eqnarray}
where $M_B$ is the Borel parameter introduced by the Borel
transformation, ${\cal F}  = (\alpha + \beta) m_c^2 - \alpha \beta
s$, ${\cal H}  = m_c^2 - \alpha (1 - \alpha) s$ and the integration
limits are given by $\alpha_{min} = (1 - \sqrt{1 - 4 m_c^2/s}) / 2$,
$\alpha_{max} = (1 + \sqrt{1 - 4 m_c^2 / s}) / 2$ and $\beta_{min} =
\alpha m_c^2 /(s \alpha - m_c^2)$.

For the $2^{+}$ tetraquark state:
\begin{eqnarray}
\rho_2^{pert}(s) \! \! \! & = & \! \! \! -\frac{1}{2^8 \pi^6}
\int_{\alpha_{min}}^{\alpha_{max}} \frac{d \alpha}{\alpha^3}
\int_{\beta_{min}}^{1 - \alpha} \frac{d \beta}{\beta^3} \bigg[
\left( 1 - \alpha - \beta \right)(\alpha + \beta) {\cal F}^4 \nonumber \\
\! \! \! & - & \! \! \! (m_u + m_d) m_c \left( 1 - (\alpha +
\beta)^2 \right) (\alpha + \beta) {\cal F}^3 \bigg]\;,
\label{rhopert} \\
\rho_2^{\langle \bar{q} q \rangle}(s) \! \! \! & = & \! \! \! \frac{
\langle \bar{q} q \rangle}{2^4 \pi^4}
\int_{\alpha_{min}}^{\alpha_{max}} \frac{d \alpha}{\alpha} \bigg[ -
\frac{(m_u + m_d) {\cal H}}{2 (1 - \alpha) \alpha} +
\int_{\beta_{min}}^{1-\alpha} \frac{d \beta}{\beta} \big[ \frac{m_c
\left(\alpha + \beta \right)^2}{\alpha \beta}
{\cal F}^2 \nonumber \\
\!\!\! & + & \!\!\! (m_u + m_d)\left({\cal F} -
2m_c^2 \right) {\cal F}\big] \bigg] \;, \\
\rho_2^{\langle g_s^2 G^2 \rangle}(s) \! \! \! & = & \! \! \!
-\frac{\langle g_s^2 G^2 \rangle}{3 \times 2^{8} \pi^6}
\int^{\alpha_{max}}_{\alpha_{min}} \frac{d \alpha}{\alpha} \bigg[
\frac{{\cal H}^2}{ 2^4 (1 - \alpha)} + \frac{1}{2^5 \alpha}
\int_{\beta_{min}}^{1-\alpha} \frac{d \beta}{\beta^2} \big[ \left( 1
- 4\alpha - 4\beta + \alpha^2 + \beta^2 \right) \nonumber
\\ \!\!\! & \times & \!\!\! {\cal F}^2 - \frac{m_c^2}{\alpha^2 \beta}
(\alpha + \beta)(\alpha \beta - \alpha^2 - \beta^2 + \alpha^3 +
\beta^3 ){\cal F}\big] \bigg] \;, \\
\rho_2^{\langle g_s \bar{q} \sigma \cdot G q \rangle}(s) \! \! \! &
= & \! \! \! - \frac{\langle g_s \bar{q} \sigma \cdot G q \rangle
m_c}{2^5 \pi^4} \int_{\alpha_{min}}^{\alpha_{max}} \frac{d
\alpha}{\alpha} \bigg[ \frac{\left( {\cal H} + m_c( m_u + m_d) (1 -
\alpha) \alpha \right)}{(1 - \alpha)} + \frac{{\cal H}}{6} \nonumber
\\ \!\!\! & + & \!\!\! \int_{\beta_{min}}^{1 - \alpha } d \beta \big[
\frac{(\alpha + \beta){\cal F}}{\alpha \beta} + \frac{1}{12} \left(
2{\cal F} - m_c \left( m_u + m_d \right) \right) \big] \bigg] \;, \label{O5} \\
\rho_2^{\langle \bar{q} q \rangle^2}(s) \! \! \! & = & \! \! \!
\frac{\langle \bar{q} q \rangle^2 }{3 \times 2^3 \pi^2}
(m_u + m_d - 4m_c)m_c \sqrt{1 - 4m_c^2/s}  \;, \label{O6} \\
\rho_2^{\langle g_s^3 G^3 \rangle}(s) \! \! \! & = & \! \! \!
\frac{\langle g_s^3 G^3 \rangle}{3 \times 2^{9} \pi^6}
\int^{\alpha_{max}}_{\alpha_{min}} d \alpha \int^{1 -
\alpha}_{\beta_{min}} \frac{d \beta}{\beta^3} (\alpha^2 + \beta^2 +
2 \alpha \beta - \alpha - \beta)(2 m_c^2 \alpha + {\cal F})\; ,
\label{O6G}
\end{eqnarray}
\begin{eqnarray}
\Pi_2^{\langle \bar{q} q \rangle^2} (M_B^2) \! \! \! & = & \! \! \!
- \frac{\langle \bar{q} q \rangle^2}{3 \times 2^3 \pi^2} (m_u + m_d)
m_c^5 \int_0^1 \frac{d \alpha}{(1 - \alpha)^2
\alpha^2} \text{Exp}[-\frac{m_c^2}{(1 - \alpha)\alpha M_B^2}]\, , \\
\Pi_2^{\langle g_s \bar{q} \sigma \cdot G q \rangle \langle \bar{q}
q \rangle} (M_B^2) \! \! \! & = & \! \! \! - \frac{\langle g_s
\bar{q} \sigma \cdot G q \rangle \langle \bar{q} q \rangle}{3 \times
2^5 \pi^2}\int^1_0 d \alpha \bigg[ \frac{2 m_c^2}{3 \alpha} -
\frac{m_c^3(m_u + m_d)}{6 M_B^2 (1 - \alpha) \alpha} + \frac{(m_u +
m_d)m_c^5}{M_B^4 (1 - \alpha)^2 \alpha^2} \nonumber \\ \!\!\! & - &
\!\!\! \frac{m_c}{(1 - \alpha)^2 \alpha^2} \left( 8 m_c \left( 1 -
\alpha \right)^2 \alpha^2  - (m_u + m_d) \left( 1 - \alpha \right)^2
\alpha^2 \right)
\nonumber \\
\!\!\! & - & \!\!\! \frac{m_c^3}{M_B^2 (1 - \alpha)^2 \alpha^2
}\left(8 m_c (1 - \alpha)\alpha - (m_u + m_d) (1 - \alpha)\alpha
\right) \bigg] \nonumber \\ \!\!\! & \times & \!\!\! \text{Exp}[-
\frac{m_c^2}{(1 - \alpha) \alpha M_B^2}] \; . \; \; \; \; \; \; \;
\; \; \; \label{rhoO8}
\end{eqnarray}

Matching the OPE side and the phenomenological side of the
correlation function $\Pi(q^2)$, i.e. the quark-hadron duality, and
performing the Borel transformation, the sum rule for the mass of
the hidden charm tetraquark state is determined to be:
\begin{eqnarray}
m_i^{c}(s_0, M_B^2) = \sqrt{- \frac{R_{i \, 1}(s_0, M_B^2)}{R_{i \,
0}(s_0, M_B^2)}} \; , \label{tetraquarkmass}
\end{eqnarray}
with
\begin{eqnarray}
R_{i \, 0}(s_0, M_B^2) & = & \int_{4 m_c^2}^{s_0} d s \;
\rho_i^{OPE}(s) e^{- s / M_B^2} + \Pi_i^{\langle O_6 \rangle}
(M_B^2) + \Pi_i^{\langle g_s \bar{q} \sigma
\cdot G q \rangle \langle \bar{q} q \rangle} (M_B^2) \; , \\
R_{i \, 1}(s_0, M_B^2) & = &
\frac{\partial}{\partial{M_B^{-2}}}{R_{i \, 0}(s_0, M_B^2)} \; ,
\end{eqnarray}
where $\langle O_6 \rangle$ represents $\langle g_s^3 G^3 \rangle$
or $\langle \bar{q} q \rangle^2$ for $i = 1$ or $i = 2$.

\section{Numerical Analysis}

In the numerical calculation, the values of the condensates and the
quark masses are used as \cite{Matheus:2006xi, Narison:2002pw,
Cui:2011fj}:
\begin{eqnarray}
\begin{aligned}
& m_u = 2.3 \; \text{MeV} \;, & & m_d = 6.4 \; \text{MeV} \; , \\
& m_c (m_c) = (1.23 \pm 0.05) \; \text{GeV} \; , & & m_b (m_b) =
(4.24 \pm 0.06) \; \text{GeV}, \\
& \langle \bar{q} q \rangle = - (0.23 \pm 0.03)^3 \; \text{GeV}^3 \;
, & & \langle g_s^2 G^2 \rangle = 0.88 \; \text{GeV}^4 \;, \\
& \langle \bar{q} g_s \sigma \cdot G q \rangle = m_0^2 \langle
\bar{q} q \rangle \; , & & \langle g_s^3 G^3 \rangle = 0.045 \;
\text{GeV}^6
\; ,\\
& m_0^2 = 0.8 \; \text{GeV}^2\;. & &
\end{aligned}
\end{eqnarray}

In the QCD Sum Rules, to select the appropriate threshold $s_0$ and the Borel parameter $M_B^2$, there are two criteria \cite{Shifman, Reinders:1984sr, P.Col}. As the convergence of the OPE must be retained in order to determine their convergence, it is essential to compare the relative contributions of each term to the total contributions on the OPE side.

The second criterion to constrain the $M_B^2$ is that the pole
contribution (PC) must be larger than the continuum contribution. Thus, for various values of the $M_B^2$, it is necessary to analyze the relative pole contribution, defined as the pole contribution divided by the total contribution, i.e. pole plus continuum. To safely eliminate the contributions of the higher excited and continuum states, the PC is generally greater than $50\%$ \cite{P.Col, Matheus:2006xi}, which is slightly different from the constraint used in \cite{Chen:2010ze}.

To find a proper value for $\sqrt{s_0}$, we carry out a similar analysis as in Refs.\cite{P.Col,Finazzo:2011he}. Since the continuum threshold is connected to the mass of the studied state by the relation $\sqrt{s_0} \sim m_i^c + 0.5 \, \text{GeV}$, various $\sqrt{s_0}$ satisfying this constraint are taken into account. Among these values, one needs then to find out the proper one which has an optimal window for Borel parameter $M_B^2$. That is, within this window, the physical quantity, here the tetraquark mass $m_i^c$, is independent of the Borel parameter $M_B^2$ as much as possible. Through the above procedure one obtains the central value of $\sqrt{s_0}$. However, in practice, in the QCD Sum Rules calculation, it is normally acceptable to vary the $\sqrt{s_0}$ by $0.1 \, \text{GeV}$ \cite{Finazzo:2011he}, which gives the lower and upper bounds and hence the uncertainties of $\sqrt{s_0}$.

\subsection{ $1^{-}$ Hidden Charm Tetraquark State}

\begin{figure}[!htb]
\begin{center}
\includegraphics[width=7.5cm]{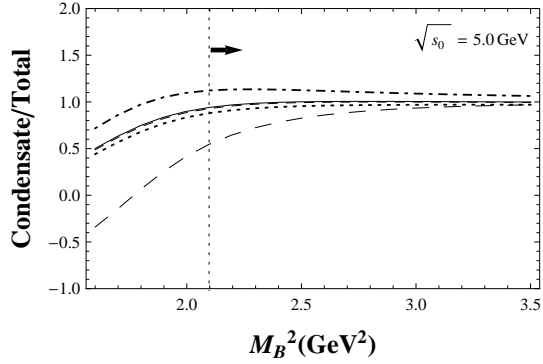}
\caption{The OPE convergence in the region $1.6 \leq M_B^2 \leq 3.5
\; \text{GeV}^2$ for the $J^P = 1^-$ hidden charm tetraquark state
with $\sqrt{s_0} = 5.0 \; \text{GeV}$. The solid line denotes the
perturbative contribution, and each subsequent line denotes the
addition of one extra condensate, {\it i.e.}, $+ \langle \bar{q} q
\rangle$ (short-dashed line), $+ \langle g_s^2 G^2 \rangle$ (dotted
line), $+ \langle g_s \bar{q} \sigma \cdot G q \rangle$ + $\langle
g_s^3 G^3 \rangle$ (dotted-dashed line), $+ \langle \bar{q} q
\rangle^2$ (long-dashed line).} \label{convergence1-}
\end{center}
\end{figure}

The OPE convergence of the $1^-$ hidden charm tetraquark state is
shown in Fig.\ref{convergence1-}, which reflects a strong OPE
convergence for $M_B^2 \geq 2.1 \; \text{GeV}^2$, making it possible
to determine the lower limit constraint of the $M_B^2$.

\begin{figure}[!htb]
\begin{center}
\includegraphics[width=7.5cm]{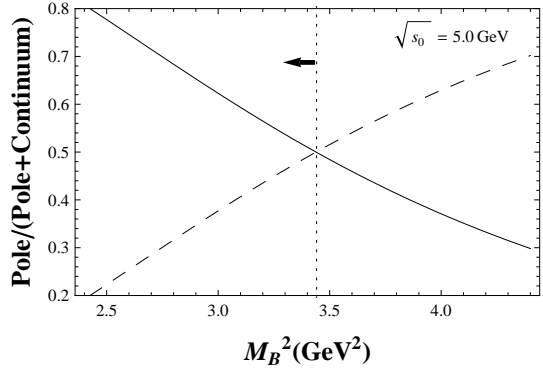}
\caption{The relative pole contribution for the $J^P = 1^-$ hidden
charm tetraquark state with $\sqrt{s_0} = 5.0 \; \text{GeV}$. The
solid line represents the relative contribution, whereas the dashed
line corresponds to the continuum contribution.} \label{pole1-}
\end{center}
\end{figure}

The result of the PC is shown in Fig.\ref{pole1-}, which indicates
the upper limit constraint of the $M_B^2$. Noting that the upper
limit constraint of the $M_B^2$ depends on the threshold value
$s_0$, for different $s_0$, there are different upper limits of the
$M_B^2$. To determine an appropriate value of the $s_0$, a similar
analysis is utilized as was applied in Ref.\cite{Matheus:2006xi,
Albuquerque:2012zy}. Thus, for the $s_0$, $\sqrt{s_0} = 5.0 \;
\text{GeV}$, the $M_B^2 \leq 3.4 \, \text{GeV}^2$.

\begin{figure}[!htb]
\begin{center}
\includegraphics[width=7.5cm]{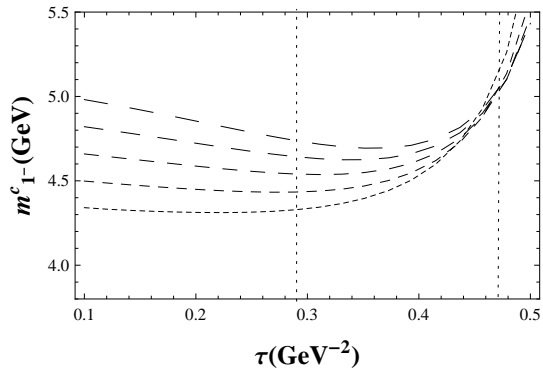}
\caption{Dependence of $m_{1^-}^c$ on the parameter $\tau$ for the
$J^P = 1^-$ hidden charm tetraquark state, where $\tau = 1 / M_B^2$,
and the continuum thresholds $\sqrt{s_0}$ are respectively taken as
$4.6$, $4.8$, $5.0$, $5.2$ and $5.4 \, \text{GeV}$, from down to up.
Two vertical lines have been placed to indicate the chosen Borel
window.} \label{mass1-}
\end{center}
\end{figure}

The dependence of $m_{1^-}^c$ is drawn on the parameter $\tau$ in
Fig.\ref{mass1-}, where $\tau = 1/ M_B^2$, and the continuum
threshold parameters $\sqrt{s_0}$ are respectively taken as $4.6$,
$4.8$, $5.0$, $5.2$, and $5.4 \, \text{GeV}$, from down to up.

The mass of the $1^-$ hidden charm tetraquark state was determined
to be:
\begin{eqnarray}
m_{1^-}^c = (4.54 \pm 0.20) \, \text{GeV} \; ,
\end{eqnarray}
where $M_B^2$ was $3.4 \; \text{GeV}^2$, and the errors stemmed from
the uncertainties of the charm quark mass, the condensates and the
threshold parameter $\sqrt{s_0}$.

\subsection{$2^+$ Hidden Charm Tetraquark State}

\begin{figure}[!htb]
\begin{center}
\includegraphics[width=7.5cm]{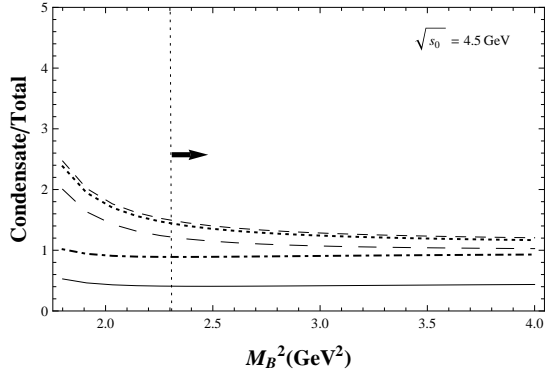}
\caption{The OPE convergence in the region $1.6 \leq M_B^2 \leq 4.0
\; \text{GeV}^2$ for the $2^+$ hidden charm tetraquark state with
$\sqrt{s_0} = 4.5 \; \text{GeV}$. The solid line denotes the
perturbative contribution, and each subsequent line denotes the
addition of one extra condensate, {\it i.e.}, $+ \langle \bar{q} q
\rangle$ (short-dashed line), $+ \langle g_s^2 G^2 \rangle$ (dotted
line), $+ \langle g_s \bar{q} \sigma \cdot G q \rangle$ + $\langle
g_s^3 G^3 \rangle$ (dotted-dashed line), $+ \langle \bar{q} q
\rangle^2$ (long-dashed line).} \label{convergence2+}
\end{center}
\end{figure}

For the $2^+$ sector, the OPE convergence is shown in
Fig.\ref{convergence2+}, which reflects a strong OPE convergence for
$M_B^2 \geq 2.3 \; \text{GeV}^2$, enabling the determination of the
lower limit constraint of the $M_B^2$.

\begin{figure}[!htb]
\begin{center}
\includegraphics[width=7.5cm]{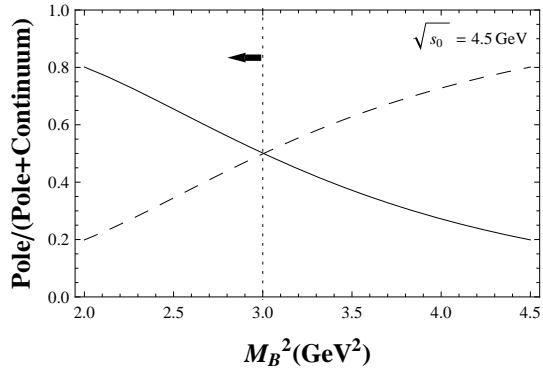}
\caption{The relative pole contribution of the the $J^P = 2^+$
hidden charm tetraquark state with $\sqrt{s_0} = 4.5 \; \text{GeV}$.
The solid line represents the relative contribution, whereas the
dashed line corresponds to the continuum contribution.}
\label{pole2+}
\end{center}
\end{figure}

The result of the PC is shown in Fig.\ref{pole2+}, which indicates
the upper limit constraint of the $M_B^2$. For the appropriate
$s_0$, $\sqrt{s_0} = 4.5 \; \text{GeV}$, the $M_B^2 \leq 3.0 \;
\text{GeV}^2$. Therefore a reliable Borel window, $2.3 \leq M_B^2
\leq 3.0 \; \text{GeV}^2$, is obtained.

\begin{figure}[!htb]
\begin{center}
\includegraphics[width=7.5cm]{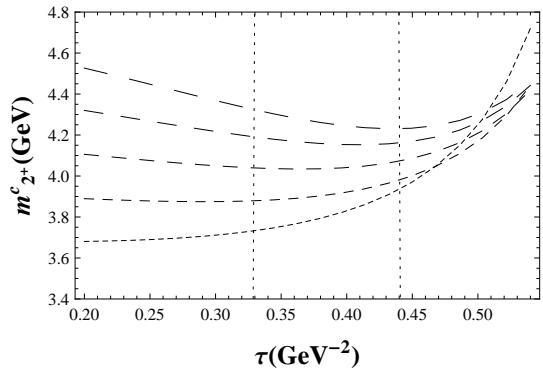}
\caption{Dependence of $m_{2^+}^c$ on the parameter $\tau$, where
$\tau = 1 / M_B^2$, and the continuum threshold parameter
$\sqrt{s_0}$ are respectively taken as $3.9$, $4.2$, $4.5$, $4.8$,
and $5.1\, \text{GeV}$, from down to up. Two vertical lines have
been placed to indicate the chosen Borel window.} \label{mass2+}
\end{center}
\end{figure}

The dependence of $m_{2^+}^c$ is drawn on the parameter $\tau$ in
Fig.\ref{mass2+}, where $\tau = 1/ M_B^2$, and the continuum
thresholds $\sqrt{s_0}$ are respectively taken as $3.9$, $4.2$,
$4.5$, $4.8$, and $5.1 \, \text{GeV}$, from down to up. In
Fig.\ref{mass2+}, the optimal mass curve of the $2^+$ hidden charm
tetraquark state is shown with $\sqrt{s_0} = 4.5 \; \text{GeV}$,
where both the aforementioned criteria are satisfied.

The mass of the $2^+$ hidden charm tetraquark was determined to be:
\begin{eqnarray}
m_{2^+}^c = (4.04 \pm 0.19) \, \text{GeV} \; ,
\end{eqnarray}
where $M_B^2$ was $3.0 \; \text{GeV}^2$, and the errors stemmed from
the uncertainties of the charm quark mass, the condensates and the
threshold parameter $\sqrt{s_0}$.

\section{Conclusions}

In this paper, we estimated the masses of the hidden charm tetraquark states with $J^P = 1^-$ and $2^+$, which are possible quantum numbers possessed by the charmonium-like resonance $Z_c(4025)$. In our calculations contributions up to dimension eight in the OPE were taken into account. Noticeably, with the $1^-$ current, the mass was found to be $m_{1^-}^c = (4.54 \pm 0.20) \; \text{GeV}$, so we deduced that such a tetraquark structure was not the candidate for $Z_c(4025)$. As in the discussions in Ref.\cite{Chen:2010ze}, it may correspond to the charged partner of the charmonium-like state $Y(4360)$ or $Y(4660)$, within the uncertainties. However, in the case of $2^+$, we found that $m_{2^+}^c = (4.04 \pm 0.19) \; \text{GeV}$, which is consistent within the errors with the experimental data of the $Z_c^+(4025)$ resonance.

As was mentioned in the introduction, the existing analyses favor $Z_c(3900)$ having quantum number of $J^P = 1^+$, and these calculations can not discriminate $Z_c(3900)$ with the $Z_c(3885)$. Possibly they may have the same origin. In this work, we calculate the masses of $J^P=2^+$ tetraquark states in the framework of QCD Sum Rules. Our result suggests that the mass of hidden charm tetraquark state is a bit more than 4 GeV, which in certain degree agrees with the recent observations of $Z_c(4025)$ or $Z_c(4020)$ by BESIII Collaboration. For the b-quark sector, by virtue of the similar numerical analysis, the masses of the tetraquark state $[b u][\bar{b}\bar{d}]$ are obtained as $m_{1^-}^b = (10.97 \pm 0.25) \; \text{GeV}$ and $m_{2^+}^b = (10.35 \pm 0.25) \; \text{GeV}$, which future experiments may verify.

Addentum: during the finalization of this work, two reports about
$Z_c(4025)$ appear, wherein the molecular picture \cite{He:2013nwa} and
initial-single-pion-emission mechanism \cite{Wang:2013qwa} were employed.

\vspace{.7cm} {\bf Acknowledgments} \vspace{.3cm}

This work was supported in part by the National Natural Science Foundation of China(NSFC) under the grants 10935012, 11121092, 11175249 and 11375200.

\end{document}